\documentclass[preprint,12pt,nonatbib]{elsarticle} 
\makeatletter
\let\c@author\relax
\makeatother

\newcommand{\Kbar}{\ensuremath{\overline{K}}}
\newcommand{\etap}{\ensuremath{\eta'}}
\newcommand{\piz}{\ensuremath{\pi^{0}}}
\newcommand{\pip}{\ensuremath{\pi^{+}}}
\newcommand{\pim}{\ensuremath{\pi^{-}}}
\newcommand{\kp}{\ensuremath{K^{+}}}
\newcommand{\km}{\ensuremath{K^{-}}}

\newcommand{\jpsi}{\ensuremath{J/\psi}}
\newcommand{\etac}{\ensuremath{\eta_{c}}}
\newcommand{\chiczero}{\ensuremath{\chi_{c0}}}
\newcommand{\chicone}{\ensuremath{\chi_{c1}}}
\newcommand{\chictwo}{\ensuremath{\chi_{c2}}}

\usepackage[backend=bibtex, sorting=none, citestyle=numeric-comp]{biblatex}

\usepackage{amssymb}
\usepackage{amsmath}
\usepackage{wrapfig}

\begin{document}

\begin{frontmatter}


\title{Mass creation by the strong interaction: \\
Glueballs - status and perspectives}

\author[RUB]{Ulrich Wiedner} 
\ead{wiedner@ep1.rub.de}

\affiliation[RUB]{organization={Ruhr-Universität Bochum},
            addressline={Universitätsstr. 150}, 
            city={Bochum},
            postcode={44801}, 
            country={Germany}}

\begin{abstract}
Glueballs represent a fascinating aspect of the strong interaction in nature. Gluons that serve as the mediators of the strong interaction are massless particles, but they possess a property unique to the strong interaction called color charge, which is analogous to electric charge in the electromagnetic interaction. Glueballs are composed of multiple gluons and would be massless without color charges. The interaction of the color charges, however, makes glueballs becoming massive objects. Glueballs thus offer a unique way to study the mass creation of strongly interacting particles.

\end{abstract}


\end{frontmatter}


\section{Introduction}
\label{sec1}

The strong interaction between gluons is characterized by an attractive force of the color charges. This interaction is evident in lattice calculations based on the QCD Lagrangian, which predict the formation of a flux tube of gluons connecting a quark and an antiquark within a meson. Furthermore, the mutual attraction between gluons suggests the possibility of forming meson-like bound states of gluons, even in the absence of quarks. Hadrons are characterized e.g. by their mass, width and dedicated quantum numbers. While the Higgs mechanism might be responsible for the masses of the elementary particles in the Standard Model, the mass-creation mechanism for hadrons is quite different. The Higgs mechanism is responsible e.g. for only a few percent of the proton mass. It is worth noting, that glueballs themselves, consisting solely of massless gluons, would be massless in the absence of the strong interaction. 

\setlength{\intextsep}{10pt}
\begin{wrapfigure}{r}{0.5\textwidth}
\centering
    \includegraphics[scale=0.40]{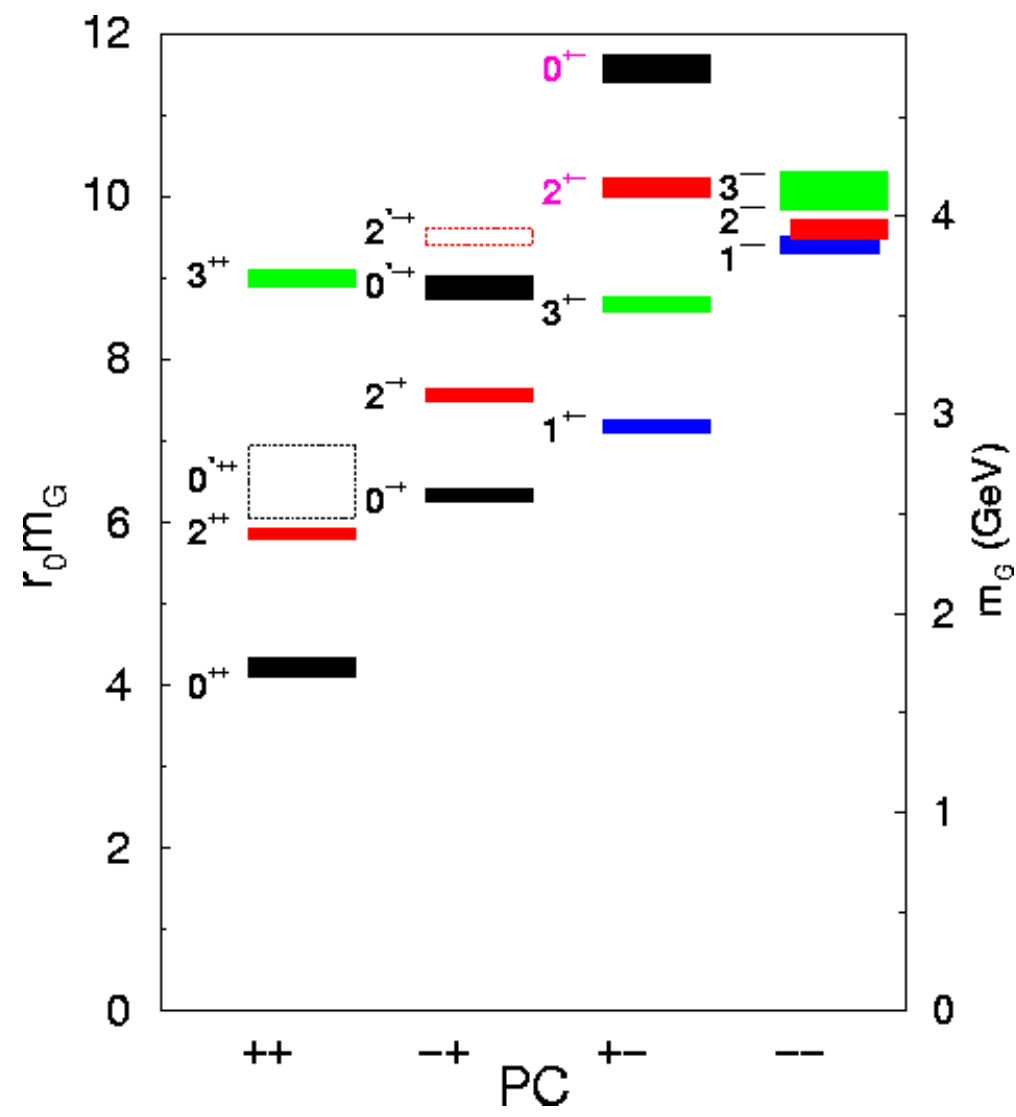}
\caption{The Yang-Mills glueball spectrum}\label{glueball}
\end{wrapfigure}

Consequently, glueballs provide a distinctive approach to investigate the mass generation of strongly interacting particles. The possibility to study the creation of massive objects from massless gauge bosons might provide also deeper insights into another poorly understood force of nature, the gravitation. Already in the Feynman lectures on Gravitation there has been mentioning of a hypothetical graviton-graviton interaction. Clearly, it is hard to see how a graviton-graviton interaction could be measured in a controlled manner due to the nature of gravitation, but glueballs do offer at least the possibility of a systematic study of the non-perturbative interaction of other massless gauge bosons beside gravitons in the laboratory. It remains to be seen if the results from glueball studies have implications for the deeper understanding of gravitational forces.

\section{The theoretical landscape}

Quenched lattice calculations provide evidence for a wide range of bound gluon states, categorized by quantum numbers and mass predictions. Figure \ref{glueball}, derived from lattice calculations \cite{Morningstar:1999rf}, offers an overview of these states. 
Superstring theory emerges as a promising candidate for a quantum theory of gravitation. Initially developed to describe hadrons, string theory encountered significant challenges when extended to a four-dimensional framework. To address these issues, the theory necessitates the incorporation of both bosonic and fermionic degrees of freedom, interconnected through the principle of supersymmetry. Furthermore, string theory necessitates a 10-dimensional space-time, which appears incompatible with the structure of the strong interaction. Despite these apparent contradictions, intriguing similarities exist between superstring theory and the strong interaction. In superstring theory, gravitation is represented by closed strings, while lattice calculations reveal glueballs as closed loops of gluons. This conceptual alignment aligns well with our research on the structure of glueballs, where we propose closed fluxes of colored gluons with twists or knots that confer distinct masses and quantum numbers to these particles \cite{Faddeev:2003aw}. If a connection between superstrings and certain aspects of the strong interaction is indeed established, a comprehensive study of glueballs holds the potential to enhance our understanding of gravitation.

The AdS/CFT (Anti-de Sitter space/conformal field theory) correspondence has emerged as a relationship between modern superstring theory and quantum chromodynamics (QCD). This correspondence establishes a mapping between a weakly coupled supergravity theory and a strongly coupled large-N gauge theory. The subsequent advancement of superstring theory models has subsequently inspired phenomenological approaches to QCD, referred to as AdS/QCD. These approaches enable the prediction of meson masses and couplings that are typically accurate to within 10\%. Predictions for glueball masses are also available and can potentially be verified. For a comprehensive review of this subject matter, please refer to the review article authored by J. Erdmenger \cite{Erdmenger:2007cm}.

\begin{wrapfigure}{hr}{0.5\textwidth}
\centering
\vspace{-1cm}
    \includegraphics[scale=0.375]{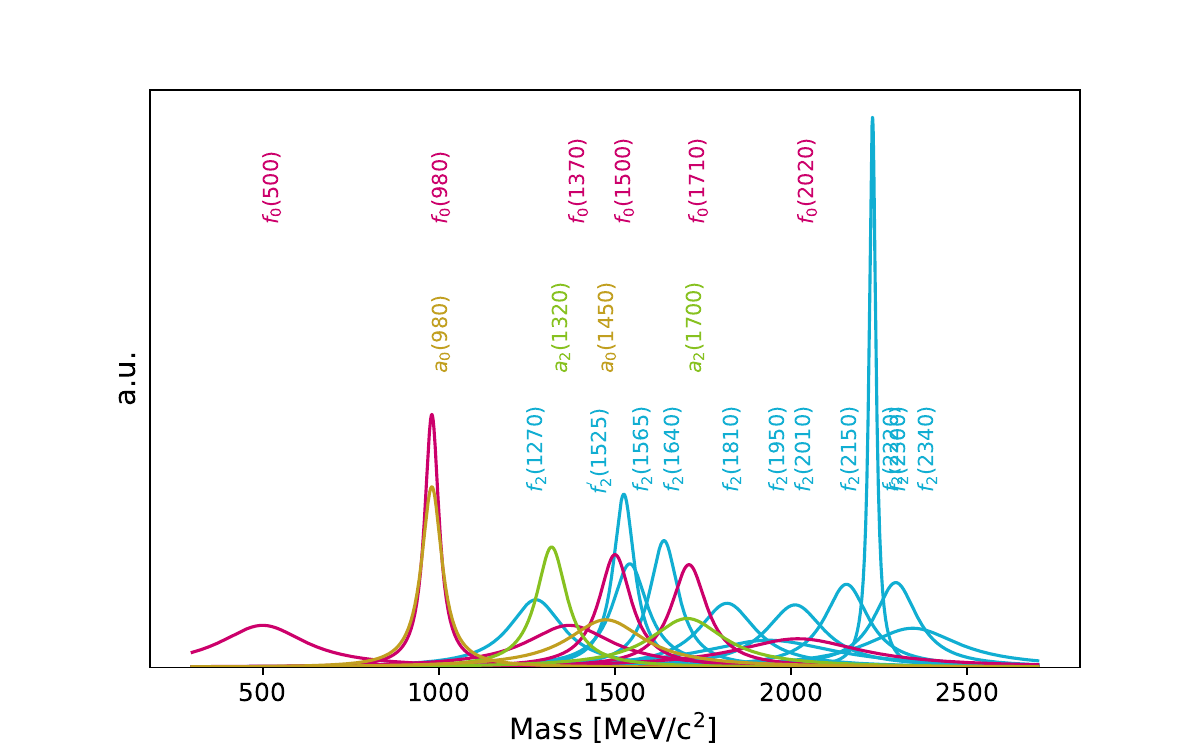}
\caption{Light hadron spectrum of $a$- and $f$-resonances based on established states according to \cite{ParticleDataGroup2024}.}\label{light_hadron}
\end{wrapfigure}

\section{The experimental status}

Does the structure of a glueball truly resemble a closed flux of color as proposed by contemporary models? How can the unknown structure and properties of glueballs be addressed and studied? Spectroscopy is as a proven approach in physics, and if feasible with multiple glueballs, it could indeed be the sole realistic avenue to enhance knowledge in this domain. Figure \ref{glueball} illustrates that the lightest glueball, as determined by lattice calculations, is the scalar glueball with the quantum numbers $J^{PC} = 0^{++}$. However, within the mass region, numerous mesons are predicted, and several possess identical quantum numbers (Fig. \ref{light_hadron}). Mixing between overlapping resonances will ensue, thereby complicating the interpretation of the glueball as a distinct entity due to its interaction with the broad light-quark states. Ideally, glueballs should manifest as additional states beyond the states of the conventional $q\bar{q}$ meson nonet. The lowest mass scalar nonet in the quark model can accommodate two mesons with quantum numbers identical to those of the scalar glueball. Nevertheless, experiments have identified four candidates: $f_0(980)$, $f_0(1370)$, $f_0(1500)$ and $f_0(1710)$ \cite{ParticleDataGroup2024}. Only two of these candidates could potentially belong to the standard meson nonet in addition to the scalar glueball with the same quantum numbers. The resulting mixing due to the overlap of the resonances further complicates its identification.

In the discussed scalar nonet mixed with the lightest scalar glueball, the $f_0(500)$ state is omitted due to its substantial width (ranging from 100 to 800 MeV) and low mass, which distinguish it from the other $f_0$ states listed above. Additionally, the $f_0$ states exceeding 2 GeV are disregarded because most candidates lack strong evidence. Also, the extensive number of candidates within a relatively narrow mass range complicates the disentanglement of their properties (refer to Fig. \ref{light_hadron}). Notably, the $f_0(980)$ state stands out among the four scalar particles as being exceptionally narrow and situated near the $K\bar{K}$ threshold. Its potential role will be elaborated further down. 

What would the decay pattern of a pure scalar glueball resemble? In the light meson sector, an investigation of the pseudoscalar nonet was conducted in reference \cite{Amsler:1995td} because no admixture of a gluonic component is anticipated in this mass region. Naively, a glueball is expected to decay flavor-blindly, resulting in branching fractions for its decay of $\pi\pi: K\Kbar:\eta\eta:\eta\etap$ $=$ $3:4:1:0$.

However, the presence of a glueball overlapping and mixing with the mesons complicates the picture and disturbs the decay pattern. The results of the mixing of the bare $\bar{q}q$ states with a glueball component can be described by a mixing matrix, which was first explored in reference \cite{Amsler:1995td} for the scalar nonet. The conclusion was that all three, $f_0(1370)$, $f_0(1500)$ and $f_0(1710)$, contain gluonic components to varying degrees due to the presence of the scalar glueball in addition to light-quark and strange-quark states.

Even though the light meson spectrum has been studied intensively by different experiments, no significant further insights on the scalar glueball ground state were achieved. Therefore, we would like to take a different approach which might also shed additional light on scalar glueballs. We propose to study the first excited scalar glueball state which has never been done before. While quenched lattice calculation put the exited $0^{++}$ glueball in the mass region below 3 GeV, a newer unquenched calculation predicts it in the mass region of the  $\chi_c$ charmonia \cite{Gregory:2012hu}. There is distinct probability that it might mix with the {\chiczero} charmonium. Such mixing was discussed in context studies of the process $e^+e^- \rightarrow J/\psi + h$ $(h = \etac$ or $\chiczero$ which showed a factor of 2 enhancement of the $\chiczero$  production compared to the {\etac} production, possibly due to the additional presence of a scalar glueball \cite{Brodsky:2003hv}. How could one establish if the \chiczero is mixed with a glueball? As mentioned above, the mixing scheme between meson nonet states and the ground state scalar glueball was developed by comparison to the unmixed pseudoscaler $J^{PC}$ = $0^{-+}$) nonet. We plan to take a similar approach of a comparison and look at {\chictwo} decays as an unmixed charmonium state and compare them to {\chiczero} decays in order to find out if there is an additional glueball component to the expected charmonium decay pattern.

\begin{table}[h]
\centering
\caption{Decay branching fractions of $\chi_{c0}$ and $\chi_{c2}$ from \cite{ParticleDataGroup2024}.}
\vspace{0.5 cm}
\begin{tabular}{lcc|lcc}
\hline\hline
 \multicolumn{3}{c|}{Hadronic B.F.}  & \multicolumn{3}{c}{Electroweak B.F.} \\
\cline{1-6}
Decay & $\chi_{c0}$ & $\chi_{c2}$ & Decay & $\chi_{c0}$ & $\chi_{c2}$ \\
\hline
$2(\pi^{+}\pi^{-})$ & 2.18 & 1.12 & $\gamma J/\psi$ & 1.41 & 19.5 \\
$\pi^{+}\pi^{-}\pi^{0}\pi^{0}$ & 3.3 & 1.86 & $\gamma\gamma$ & $2.06\times10^{-2}$ & $2.91\times10^{-2}$ \\
$\pi^{+}\pi^{-}K^{+}K^{-}$ & 1.81 & 0.84 & $e^{+}e^{-}J/\psi$ & $1.34\times10^{-2}$ & $22.0\times10^{-2}$ \\
$3(\pi^{+}\pi^{-})$ & 1.96 & 1.53 & $\mu^{+}\mu^{-}J/\psi$ & $0.19\times10^{-2}$ & $2.07\times10^{-2}$ \\
$K^{+}\pi^{-}K^{0}_{S}\pi^{0} + \mathrm{cc}$ & 2.5 & 1.4 &  &  &  \\
\hline\hline
\end{tabular}
\end{table}
\vspace{0.25 cm}

What further experimental evidence exists for the mixing of the {\chiczero} with an excited scalar glueball. First of all it is an interesting observation that the width of the {\chiczero} with 10.5 MeV is significantly large than those of the {\chicone} with 0.84 MeV and the {\chictwo} with 1.97 MeV. A larger width could point to a mixing which opens additional decay channels. Glueballs decay in first order into hadrons via the strong interaction. A comparison of the most hadronic branching fractions ($2(\pip \pim)$, $\pip \pim \piz \piz$, $\pip \pim \kp \km$, $3(\pip \pim)$, $\kp \pim \overline{K}^0 \piz + cc$ \cite{ParticleDataGroup2024}) of {\chiczero} decays relative to {\chictwo} decays gives a ratio of 1.74 in favor of {\chiczero} decays. On the other hand, a gluonic component would be suppressed in its electromagnetic and weak decays. Only 7.2\% of the radiative {\chiczero} decay into $\gamma J\psi$ compared to the same decay of the {\chictwo}. A similar pattern appears for decays into $e^+e^- \jpsi$ with 6.1\% and $\mu^+\mu^- \jpsi$ with 9.2\%. The branching fractions of the most prominent hadronic decays relative to the decay into $2\gamma$ is bigger for the {\chiczero} by a factor of 1.95 relative to the {\chictwo}. Finally, the branching fraction ratio for the {\chiczero} decay into $2\gamma$ is suppressed compared to the {\chictwo} with a value of 0.71. The numbers are summarized in Tab. 1.

Despite the fact that $\chi_{c0}$ decays might reveal the excited scalar glueball, there is the distinct possibility that the decay of the $\chi_{c0}$ might help to shed light on the ground-state scalar glueball. The possible glue content of the light $f_0$ resonances is spread out among several particles due to the above mentioned mixing. In holographic inspired models for glueballs (see e.g. \cite{Faddeev:2003aw}, \cite{Sonnenschein:2015zaa}), glueballs decay by breaking the string. Such a string breaking together with the production of a $q\bar{q}$ pair at the open ends leads to the production of mesons in a glueball decay. However, it is also conceivable that an open broken string reconnects, leading to a lower mass glueball, if this is feasible from the quantum numbers. The different branching fractions might mainly reflect the main $q\bar{q}$ component appearing after the string breaking of the particles.

\setlength{\intextsep}{0pt}
\begin{wrapfigure}{hr}{0.5\textwidth}
\centering
    \includegraphics[scale=0.14]{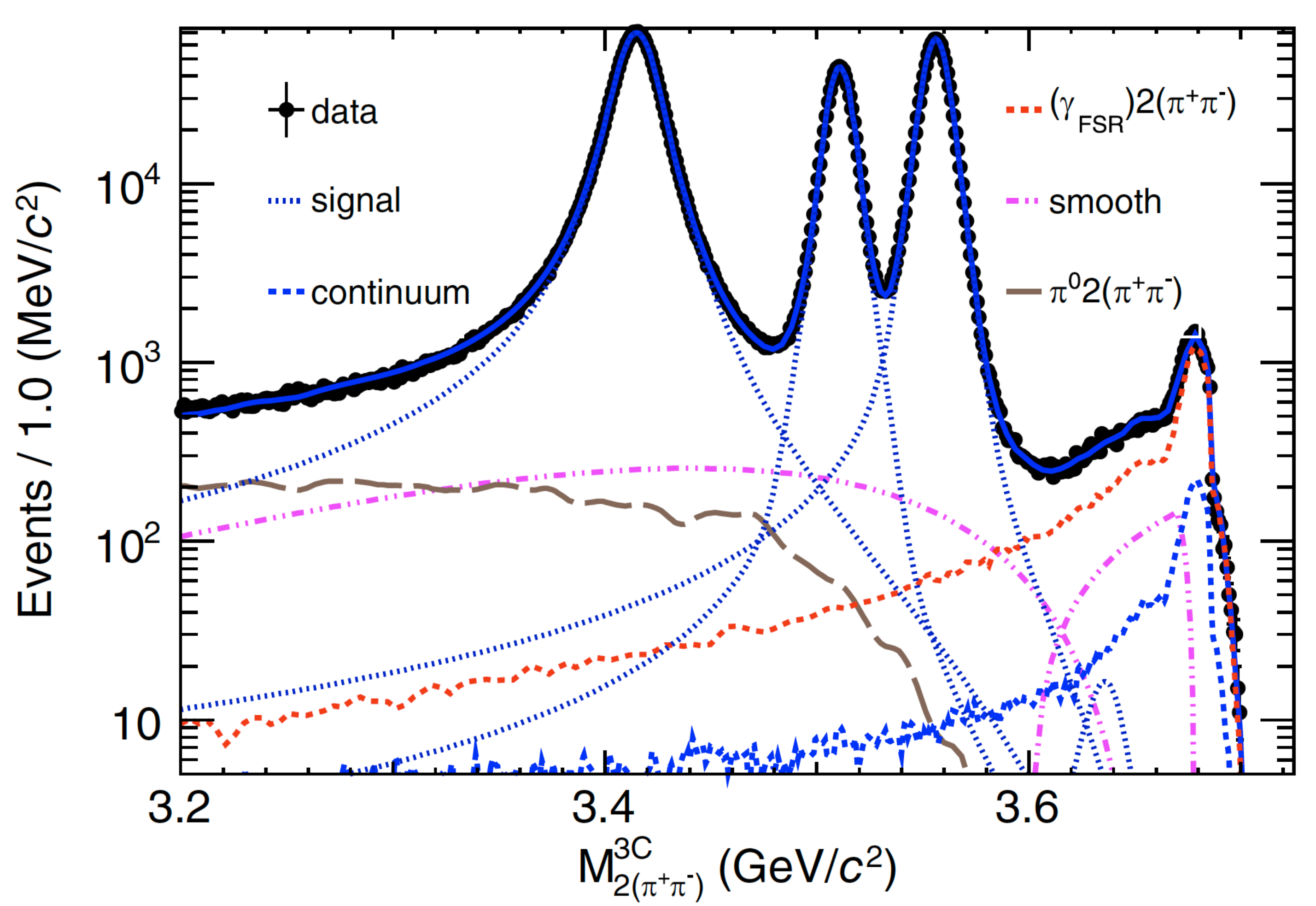}
\caption{The invariant mass distribution of $2(\pi^+\pi^-)$ from \cite{BESIII:2024jmr}. Black dots are data, the blue solid line is the total fit result. The blue dotted lines are the $\eta_c(2S)$ and $\chi_{cJ}$ signal shapes. The other lines represent different background contributions from $\psi(3686) \rightarrow \pi^0 2(\pi^+\pi^-)$ (brown long-dashed), $\psi(3686) \rightarrow (\gamma_{\text{FSR}}) 2(\pi^+\pi^-)$ (red dashed), continuum process (blue dashed-dotted line), other smooth background components (magenta dash-dot-doted).}\label{Chic_states_pict}.
\end{wrapfigure}

In case of a glueball component in the $\chi_{c0}$, a production of the $f_0$ particles containing the most glue content might be enhanced In such case, more clarity about the scalar glueball and its mixing pattern with nonet mesons will be achieved. The lightest scalar particle appearing as $f_0$ resonances possess different hadronic decay patterns into ``stable'' particles like $\pi$, $K$ and $\eta$.

We propose to select out of the high-statistics data of $2.26 \cdot 10^9$ $\psi(2S)$ events from the BESIII experiment \cite{BESIII:2024jmr} the radiative decays into $\chi_{c0}$ and $\chi_{c2}$ resonances, which represent $(9.75 \pm 0.22)\%$ and $(9.38 \pm 0.23)\%$, respectively, of all $\psi(2S)$ decays. More than $2 $ million $\chi_{c0}$ and $\chi_{c2}$ are available for analyses into different final states. The lightest scalar particle appearing as $f_0$ resonances possess different decay patterns into ``stable'' particles like $\pi$, $K$ and $\eta$. 

In order to identify and get an overall clear picture of their production branching fraction from $\chi_{c0}$ and $\chi_{c2}$ decays we plan to analyze the following final states: $K^+K^-K_SK_S$, $K^+K^-\pi^+\pi^-$, $K^+K^-\pi^0\pi^0$, $\pi^+\pi^-\pi^+\pi^-$, $\pi^+\pi^-\pi^0\pi^0$, $\pi^0\pi^0\pi^0\pi^0$, $\pi^+\pi^-\eta\eta$, $\pi^+\pi^-\eta\eta'$. By doing so, the main hadronic decay channels of the scalar $f_0$ resonances are covered and the different combinations of $f_0$ and $f_2$ states in the decay of the $\chi_{c0}$ and $\chi_{c2}$ can be determined. How clean and well-analyzable the data set from BESIII is can be seen from a different analysis in Fig. \ref{Chic_states_pict} \cite{BESIII:2024jmr}.

Back to the role of the $f_0(980)$. Based on sum rules a paper by Ellis, Fujii and Kharzeev \cite{Ellis:1999kv} discuss the possibility that the $f_0(980)$ might be a glueball candidate which is mixed with ordinary $\bar{q}q$ mesons. To verify such a scenario, however, it has to be verified with a reasonable mixing scheme with the $\bar{q}q$ mesons. Let us consider for a moment if such a glueball component could appear. That could mean that the string breaking in an excited scalar glueball leads to the production of two glueballs by reconnecting the two open strings into two closed strings again. Those might have different masses and the possibility could exist that the $f_0(980)$, mixed with $\bar{q}q$ states would be a particle containing glue and having the lower mass. Again, decay patterns of an excited scalar glueball, mixed with the $\chi_{c0}$ would help to clarify such a hypothetical picture. Therefore it might be useful to study which combinations of $f_0$ mesons appear most often and compare their decay patterns. In case of the existence of a two-glueball decay, the candidates in such a mixing scheme with the most glue component would appear most coupled together.

\section{Summary and Conclusions}\label{sec:sumandconc}

To summarize, we would like to study the decay of the $\chi_{c0}$ into $f_0$ states and compare it to $\chi_{c2}$ decays. We believe that there is enough evidence for our suspicion the $\chi_{c0}$ is mixed in with the excited scalar glueball. An excited scalar glueball will then open the possibility to shed further light on the ground state scalar glueball and glueball decays in general.


\printbibliography

\end{document}